\documentstyle[11pt,amssymb]{article}

\def\dalemb#1#2{{\vbox{\hrule height .#2pt
        \hbox{\vrule width.#2pt height#1pt \kern#1pt
                \vrule width.#2pt}
        \hrule height.#2pt}}}

\def\0{{\sst{(0)}}}
\def\1{{\sst{(1)}}}
\def\2{{\sst{(2)}}}
\def\3{{\sst{(3)}}}
\def\4{{\sst{(4)}}}
\def\5{{\sst{(5)}}}
\def\6{{\sst{(6)}}}
\def\7{{\sst{(7)}}}
\def\8{{\sst{(8)}}}

\def\tY{\widetilde Y}

\def\ep{\epsilon}
\def\td{\tilde}
\def\wtd{\widetilde}

\let\a=\alpha

\def\nn{\nonumber} \def\bd{\begin{document}} \def\ed{\end{document}}
\def\ds{\documentstyle} \let\fr=\frac \let\bl=\bigl \let\br=\bigr
\let\Br=\Bigr \let\Bl=\Bigl 
\let\bm=\bibitem
\let\na=\nabla
\let\pa=\partial \let\ov=\overline 
\newcommand{\be}{\begin{equation}} 
\newcommand{\ee}{\end{equation}} 
\def\ba{\begin{array}}
\def\ea{\end{array}}
\def\ft#1#2{{\textstyle{{\scriptstyle #1}\over {\scriptstyle #2}}}}
\def\fft#1#2{{#1 \over #2}}
\def\del{\partial}
\def\sst#1{{\scriptscriptstyle #1}}
\def\oneone{\rlap 1\mkern4mu{\rm l}}
\def\ie{{\it i.e.\ }}
\def\etc{{\it etc.\ }}
\def\via{{\it via}}
\def\semi{{\ltimes}}
\def\cv{{\cal V}}
\def\str{{\rm str}}
\def\jm{{\rm j}}
\def\im{{\rm i}}
\def\vp{{\varphi}}
\def\sech{{{\rm sech\,}}}

\def\cramp{\medmuskip = 2mu plus 1mu minus 2mu}
\def\cramper{\medmuskip = 2mu plus 1mu minus 2mu}
\def\crampest{\medmuskip = 1mu plus 1mu minus 1mu}
\def\uncramp{\medmuskip = 4mu plus 2mu minus 4mu}

\def\mapright#1{\smash{\mathop{-\!\!\!-\!\!\!-\!\!\!-\!\!\!-\!\!\!
             \longrightarrow}\limits^{#1}}}
\def\maprightt#1#2{\smash{\mathop{-\!\!\!-\!\!\!-\!\!\!-\!\!\!-\!\!\!
             \longrightarrow}\limits^{#1}_{#2}}}

\def\tX{{{\wtd X}}}
\newcommand{\ho}[1]{$\, ^{#1}$}
\newcommand{\hoch}[1]{$\, ^{#1}$}
\newcommand{\bea}{\begin{eqnarray}} 
\newcommand{\eea}{\end{eqnarray}} 
\newcommand{\ra}{\rightarrow}
\newcommand{\lra}{\longrightarrow}
\newcommand{\Lra}{\Leftrightarrow}
\newcommand{\ap}{\alpha^\prime}
\newcommand{\bp}{\tilde \beta^\prime}
\newcommand{\tr}{{\rm tr} }
\newcommand{\Tr}{{\rm Tr} } 
\newcommand{\NP}{Nucl. Phys. }
\newcommand{\tamphys}{\it Center for Theoretical Physics\\
Texas A\&M University, College Station, Texas 77843}
\newcommand{\upenn}{\it Department of Physics and Astronomy\\
University of Pennsylvania, Philadelphia, Pennsylvania 19104}

\newcommand{\auth}{M. Cveti\v{c}\hoch{\dagger1}, 
H. L\"u\hoch{\dagger1} and C.N. Pope\hoch{\ddagger2}
 }

\begin{document}
\begin{flushright}
\hfill{CTP TAMU-06/00}\\
\hfill{UPR-877-T}\\
\hfill{hep-th/0002099}\\
\hfill{February 2000}\\
\end{flushright}

\vspace{15pt}

\begin{center}
{ \large {\bf Geometry of The Embedding of Supergravity Scalar Manifolds in 
$D=11$ and $D=10$}}

\vspace{10pt}
\auth

\vspace{10pt}

{\hoch{\dagger}\upenn}

\vspace{10pt}
{\hoch{\ddagger}\tamphys}

\vspace{30pt}

\underline{ABSTRACT}
\end{center}

    Several recent papers have made considerable progress in proving
the existence of remarkable consistent Kaluza-Klein sphere reductions
of $D=10$ and $D=11$ supergravities, to give gauged supergravities in
lower dimensions.  A proof of the consistency of the full gauged
$SO(8)$ reduction on $S^7$ from $D=11$ was given many years ago, but
from a practical viewpoint a reduction to a smaller subset of the
fields can be more manageable and explicit, for the purposes of
lifting lower-dimensional solutions back to the higher dimension.  The
major complexity of the spherical reduction Ans\"atze comes from the
spin-0 fields, and of these, it is the pseudoscalars that are the most
difficult to handle.  In this paper we address this problem in two
cases.  One arises in a truncation of $SO(8)$ gauged supergravity in
four dimensions to $U(1)^4$, where there are three pairs of dilatons
and axions in the scalar sector.  The other example involves the
truncation of $SO(6)$ gauged supergravity in $D=5$ to a subsector
containing a scalar and a pseudoscalar field, with a potential that
admits a second supersymmetric vacuum aside from the
maximally-supersymmetric one.  We briefly discuss the use of these
embedding Ans\"atze for the lifting of solutions back to the higher
dimension.

{\vfill\leftline{}\vfill
\vskip 5pt
\footnoterule
{\footnotesize \hoch{1} Research supported in part by DOE grant 
DE-FG02-95ER40893 \vskip -12pt} \vskip 14pt
{\footnotesize  \hoch{2} Research supported in part by DOE 
grant DE-FG03-95ER40917.\vskip  -12pt}}

\pagebreak
\setcounter{page}{1}

\section{Introduction}

    In this paper, we discuss two examples where non-trivial subsets
of the scalar sectors of gauged supergravities are obtained by
spherical reduction from a higher dimension. The first example is the
embedding of the scalars in the $U(1)^4$ maximal abelian truncation of
$SO(8)$ gauged $N=8$ maximal supergravity in $D=4$, arising from
$D=11$ {\it via} compactification on $S^7$.  The consistency of the
full $SO(8)$ reduction on $S^7$ was proven in \cite{dwn}, although at
a somewhat implicit level.  The $N=2$ truncation
includes a total of six scalar fields, comprising three dilaton/axion
pairs.  In terms of the original $SO(8)$ representations of the full
theory, where there are 35 scalars in the $35_v$, and 35 pseudoscalars
in the $35_c$ of $SO(8)$, the three dilatons come from the $35_v$, and
the three axions come from the $35_c$.  In \cite{ten}, a further
simplifying truncation was performed, in which the three axions were
set to zero.  The reduction Ansatz becomes considerably more
complicated when axions are included, as was already seen in the case
of the single dilaton/axion pair of the $N=4$ gauged $SO(4)$
truncation, discussed in \cite{d4gauge}.  In the present example, the
inclusion of the three axions as well as the three dilatons leads to a
considerably more complicated structure in the reduction Ansatz.

   The second example is a truncation of the $SO(6)$ gauged $N=8$
maximal supergravity in $D=5$, arising from type IIB {\it via}
compactification on $S^5$.  In this case there are 42 spin-0 fields in
total, comprising 20 scalars in the $20'$ of $SO(6)$, 20 pseudoscalars
in the $10$ and $\overline{10}$, and two singlets corresponding to the
original dilaton and axion of the type IIB theory.  The truncation we
shall consider retains two spin-0 fields, comprising one scalar from
the $20'$, and one pseudoscalar from the $10$ and $\overline{10}$.
This particular truncation is of interest because it is large enough
to include the fields that participate in two distinct supersymmetric
vacua of the $D=5$ gauged theory \cite{kpw}, one with maximal $N=8$,
$SO(6)$ symmetry, and the other with $N=2$, $SU(1)\times U(1)$
symmetry.  Although an explicit interpolating solution is not known it
is in principle describable within the truncation we are making.

   In both of our examples, we shall concentrate on elucidating the
geometrical structure of the embedding in $D=11$ or type IIB
supergravity.  Specifically, we shall concentrate on the Ansatz for
the Kaluza-Klein reduction of the metric tensor.  Strictly speaking,
one can only be sure that the reduction is fully consistent with all
the equations of motion of the higher-dimensional theory if one has
the complete Ansatz for all the higher-dimensional fields, including
the antisymmetric tensor field strengths.  (Or, alternatively, if an
``existence proof'' for the consistency of the reduction Ansatz has
independently been constructed.)  Obtaining the Ansatz for the
antisymmetric tensor fields is notoriously difficult, and we shall not
complete this part of the analysis in this paper.  In the case of our
$D=4$ example, we can appeal to the results of \cite{dwn}, in which a
complete proof of the consistency of the $S^7$ reduction is exhibited.
In principle it allows one to read off the Ansatz for the 4-form field
strength, although only an implicit procedure for its construction is
presented.  On the other hand, the general Anstaz for the metric tensor {\it
is} rather explicit, and it is by making use of this expression that we are
able to obtain the $D=4$ results in this paper.   These results can be
used in order to study the eleven-dimensional geometrical structure of
general domain-wall solutions in $D=4$ supported both by the three
dilatonic scalars and also the three accompanying axions.  Such
solutions can be constructed from the purely dilatonic ones by means
of $SL(2,R)$ transformations.  

    In $D=5$ the situation is less clear, since no proof for
the consistency of the full $S^5$ reduction to $SO(6)$ gauged $N=8$
maximal supergravity currently exists.  A conjecture for the metric
reduction Ansatz appears in \cite{kpw}, which is closely analogous to
the known construction in $D=4$ given in \cite{dwn}, and it is this
that we use in order to obtain an explicit expression for the metric
embedding for our 2-scalar truncation.  Again, the complexities of the
antisymmetric tensor embedding have prevented us from obtaining a full
non-linear result in that sector.  Thus the status of our $D=5$
embedding is that, subject to the assumption of an ultimate
consistency of the $S^5$ reduction scheme,\footnote{Further evidence
for the consistency of the $S^5$ reduction was obtained in
\cite{clps}, where certain scalar plus gravity truncations in
Kaluza-Klein sphere reductions were proved to be consistent.
Additionally, the complete consistent reductions of $D=11$ supergravity
on $S^4$ \cite{vann1,vann2} and massive type IIA supergravity on $S^4$
\cite{d6gauge} have been constructed.  Recently, more
evidence for the consistency of the $S^5$ reduction was presented in
\cite{nasvam}.} and subject to the
assumption that the conjecture for the metric Ansatz in \cite{kpw} is
correct, then our explicit results for the 2-scalar metric Ansatz is
valid.  In principle, our result can then be used to study the
geometry of the RG flow describing the transition between the two
supersymmetric extrema of the associated scalar potential. 

\section{$N=2$ $U(1)^4$ Gauged Supergravity in $D=4$ From
$D=11$}

\subsection{The Three Dilaton/Axion Pairs in $D=4$}

     The $35_v + 35_c$ of spin-0 fields in $SO(8)$ gauged supergravity
in $D=4$ are described in terms of a 56-vielbein ${\cal V}$, with the
block-diagonal form
\be
{\cal V} = \pmatrix{ u_{ij}{}^{IJ} & v_{ij KL} \cr
                     v^{k\ell IJ} & u^{k\ell}{}_{KL} }\,,
\ee
which transforms under local $SU(8)$ and rigid $E_7$
\cite{dwn00,dwn0}.  In terms of the quantities $u_{ij}{}^{IJ}$ and
$v_{ij KL}$, it was shown in \cite{dwn} (having been previously
proposed in \cite{dwnw}) that the Ansatz for the inverse of the
internal $S^7$ compactifying metric is
\be
\hat g^{mn}(x,y)\equiv 
\hat\Delta^{-1}\, g^{mn}(x,y) = \ft12
(K^{m IJ}\, K^{n KL} + K^{n IJ}\, K^{m KL})\, (u_{ij}{}^{IJ} 
+ v_{ij IJ})\, (u^{ij}{}_{KL} + v^{ijKL})\,,\label{dwnic}
\ee
where $K^{m IJ}$ are the 28 Killing vectors on the round $S^7$, and
\be
\hat\Delta^2 = \fft{\det(g_{mn}(x,y))}{\det(g_{mn}(y))}\,,\label{delhat}
\ee
where $g_{mn}(x,y)$ is the inverse of $g^{mn}(x,y)$, and
$g_{mn}(y)$ is $g_{mn}(x,y)$ with the scalar fields all set to
zero, so that it becomes the round $S^7$ metric. 
The eleven-dimensional metric Ansatz will be given by \cite{dwnw,dwn}
\be
d\hat s_{11}^2  = \hat\Delta^{-1}\, ds_4^2 + g_{mn}(x,y)\, dy^m\,
dy^n   = \hat\Delta^{-1}\, (ds_4^2 + \hat g_{mn}(x,y)\, dy^m\, dy^n)
\,,\label{d11met}
\ee
where $\hat g_{mn}(x,y)=\hat\Delta\, g_{mn(x,y)}$ 
is the inverse of $\hat g^{mn}(x,y)$.\footnote{For
now, we shall leave out the Kaluza-Klein 
gauge fields from the construction of the metric.  As discussed in
\cite{duliu,ten}, the truncation to three dilaton/axion pairs is
naturally accompanied by the four $U(1)$ gauge fields of the maximal
abelian $U(1)^4$ subgroup of $SO(8)$.  These gauge fields are easily
incorporated in the Kaluza-Klein Ansatz, and we shall add them in at
the end of the derivation.  We shall also set the gauge coupling
constant $g$ equal to 1 for now, and restore it later.}

    We use the parameterisation of the $u_{ij}{}^{IJ}$ and $v_{ij KL}$
matrices described in \cite{duliu}.  In particular, we introduce three
scalars $\lambda_i$, and three associated pseudoscalars $\sigma_i$,
whose kinetic Lagrangian is
\be
{\cal L}= -\ft12 \sum_i \Big( (\del\lambda_i)^2  + \sinh^2\lambda_i \,
(\del\sigma_i)^2 \Big)\,.
\ee
To shorten the subsequent formulae, we make the following definitions:
\be
c_i \equiv \cosh\lambda_i\,,\qquad s_i \equiv \sinh \lambda_i\,.
\ee
Also, for future convenience, we introduce the ``standard''
dilaton/axion pairs $(\vp_i, \chi_i)$, related to $(\lambda_i,
\sigma_i)$ by
\bea
\cosh\lambda_i &=& \cosh\varphi_i + \ft12 \chi_i^2\, e^{\varphi_i}\,,\nn\\
\cos\sigma_i \,\sinh\lambda_i &=& \sinh\varphi_i - \ft12 \chi_i^2\,
e^{\varphi_i}\,,\label{siglam}\\
\sin\sigma_i \,\sinh\lambda_i &=& \chi_i\, e^{\varphi_i}\,.\nn
\eea
In terms of these fields, the scalar kinetic terms are
\be
{\cal L} = -\ft12 \sum_i \Big( (\del\vp_i)^2 + e^{2\vp_i}\,
(\del\chi_i)^2\Big)\,.
\ee

    After some algebra, we find that $u_{ij}{}^{IJ}$ and $v_{ij KL}$
are given by
\bea
&&\ft14 u_{ij}{}^{KL} P_{ij}\, Q_{KL} =
c_1\, (P_{a_1 a_2}\, Q_{a_1 a_2} + P_{a_3 a_4}\, Q_{a_3 a_4})
\label{uexpr}\\
&&+c_2\, (P_{a_1 a_3}\, Q_{a_1 a_3} + P_{a_2 a_4}\, Q_{a_2 a_4})
+c_3\, (P_{a_1 a_4}\, Q_{a_1 a_4} + P_{a_2 a_3}\, Q_{a_2 a_3})\nn\\
&&+ P_{12}\, (c_1 \, c_2\, c_3 \, Q_{12} + c_1\, s_2\, s_3\, 
e^{ \im(\sigma_2+\sigma_3)}\, Q_{34} + c_2\, s_1\, s_3\,
e^{\im(\sigma_1+\sigma_3)}\, Q_{56} + c_3\, s_1\, s_2\,
e^{\im(\sigma_1 + \sigma_2)}\, Q_{78})\nn\\
&&+ P_{34}\, (c_1 \, c_2\, c_3 \, Q_{34} + c_1\, s_2\, s_3\, 
e^{ -\im(\sigma_2+\sigma_3)}\, Q_{12} + c_2\, s_1\, s_3\,
e^{\im(\sigma_1-\sigma_3)}\, Q_{78} + c_3\, s_1\, s_2\,
e^{\im(\sigma_1 - \sigma_2)}\, Q_{56})\nn\\
&&+ P_{56}\, (c_1 \, c_2\, c_3 \, Q_{56} + c_1\, s_2\, s_3\, 
e^{ \im(\sigma_2-\sigma_3)}\, Q_{78} + c_2\, s_1\, s_3\,
e^{-\im(\sigma_1+\sigma_3)}\, Q_{12} + c_3\, s_1\, s_2\,
e^{\im(-\sigma_1 + \sigma_2)}\, Q_{34})\nn\\
&&+ P_{78}\, (c_1 \, c_2\, c_3 \, Q_{78} + c_1\, s_2\, s_3\, 
e^{ \im(-\sigma_2+\sigma_3)}\, Q_{56} + c_2\, s_1\, s_3\,
e^{\im(-\sigma_1+\sigma_3)}\, Q_{34} + c_3\, s_1\, s_2\,
e^{-\im(\sigma_1 + \sigma_2)}\, Q_{12})\nn\\
&&\ft14 v_{ij KL} P_{ij}\, Q_{KL} =\label{vexpr}\\
&&-s_1\, ( e^{\im\sigma_1}\, \ep^{a_1 b_1}\, \ep^{a_2 b_2}\, 
P_{a_1 a_2}\, Q_{b_1 b_2} + 
e^{-\im\sigma_1}\, \ep^{a_3 b_3}\, \ep^{a_4 b_4}\,
P_{a_3 a_4}\, Q_{b_3 b_4})\nn\\
&&-s_2\, ( e^{\im\sigma_2}\, \ep^{a_1 b_1}\, \ep^{a_3 b_3}\, 
P_{a_1 a_3}\, Q_{b_1 b_3} + 
e^{-\im\sigma_2}\, \ep^{a_2 b_2}\, \ep^{a_4 b_4}\,
P_{a_2 a_4}\, Q_{b_2 b_4})\nn\\
&&-s_3\, ( e^{\im\sigma_3}\, \ep^{a_1 b_1}\, \ep^{a_4 b_4}\, 
P_{a_1 a_4}\, Q_{b_1 b_4} + 
e^{-\im\sigma_3}\, \ep^{a_2 b_2}\, \ep^{a_3 b_3}\,
P_{a_2 a_3}\, Q_{b_2 b_3})\nn\\
&&+ P_{12}\, (s_1 \, s_2\, s_3 \,
e^{\im(\sigma_1+\sigma_2+\sigma_3)}\, Q_{12} + s_1\, c_2\, c_3\, 
e^{ \im\sigma_1}\, Q_{34} + s_2\, c_1\, c_3\,
e^{\im\sigma_2}\, Q_{56} + s_3\, c_1\, c_2\,
e^{\im\sigma_3}\, Q_{78})\nn\\
&&+ P_{34}\, (s_1 \, s_2\, s_3 \,
e^{\im(\sigma_1-\sigma_2-\sigma_3)}\, Q_{34} + s_1\, c_2\, c_3\, 
e^{ \im\sigma_1}\, Q_{12} + s_2\, c_1\, c_3\,
e^{-\im\sigma_2}\, Q_{78} + s_3\, c_1\, c_2\,
e^{-\im\sigma_3}\, Q_{56})\nn\\
&&+ P_{56}\, (s_1 \, s_2\, s_3 \,
e^{\im(-\sigma_1+\sigma_2-\sigma_3)}\, Q_{56} + s_1\, c_2\, c_3\, 
e^{ -\im\sigma_1}\, Q_{78} + s_2\, c_1\, c_3\,
e^{\im\sigma_2}\, Q_{12} + s_3\, c_1\, c_2\,
e^{-\im\sigma_3}\, Q_{34})\nn\\
&&+ P_{78}\, (s_1 \, s_2\, s_3 \,
e^{\im(-\sigma_1-\sigma_2+\sigma_3)}\, Q_{78} + s_1\, c_2\, c_3\, 
e^{ -\im\sigma_1}\, Q_{56} + s_2\, c_1\, c_3\,
e^{-\im\sigma_2}\, Q_{34} + s_3\, c_1\, c_2\,
e^{\im\sigma_3}\, Q_{12})\,.\nn
\eea
Here, we have introduced $P$ and $Q$ simply as arbitrary antisymmetric
tensors, in order to provide a compact way of summarising all the
components of the $u_{ij}{}^{IJ}$ and $v_{ij KL}$ matrices.  The index
notation is as follows.  Indices with a ``1'' subscript, such as
$a_1$, range over the values $(1,2)$; similarly $a_2$ ranges over
$(3,4)$, $a_3$ ranges over $(5,6)$ and $a_4$ ranges over $(7,8)$.   

   Next, we substitute these results into the Ansatz (\ref{dwnic}) for
the inverse $S^7$ metric.  It is advantageous to introduce a new
parameterisation for the dilaton/axion pairs, as follows:
\be
Y_i\equiv  e^{\fft12\varphi_i}\,, \qquad\tY_i \equiv (1 + \chi_i^2\,
Y_i^4)^{\fft12}\, Y_i^{-1}\,, \qquad b_i\equiv \chi_i\, Y_i^2\,,
\ee
and so
\bea
\cosh\lambda_i &=& \ft12 (Y_i^2+ \tY_i^2)\,,\nn\\
\cos\sigma_i \,\sinh\lambda_i &=& \ft12 (Y_i^2- \tY_i^2)\,,\label{siglam2}\\
\sin\sigma_i \,\sinh\lambda_i &=& b_i\,.\nn
\eea
It is also advantageous to redefine the $SO(8)$ basis relative to the
one we have used so far.  The action of transformation, which amounts
to a triality rotation under which $K_{ij}\longrightarrow \ft12
(\Gamma_{ij})^{k\ell}\, K_{k\ell}$, is given explicitly in Appendix A.
After doing this, we find that the inverse internal metric
(\ref{dwnic}) takes the form\footnote{The notation for writing the
inverse metric is $\del_s^2 \equiv g^{mn}\, \del_m\, \del_n$.  The
derivatives do not act on any other objects here; it is just a
convenient way of writing all the components of $g^{mn}$ in one
formula, exactly analogous to writing the downstairs metric as $ds^2 =
g_{mn}\, dy^m\, dy^n$.  For example, the inverse of the 2-sphere
metric $ds^2 = d\theta^2 + \sin^2\theta\, d\phi^2$ is written as
\be
\del_s^2 = \del_\theta^2 + \fft{1}{\sin^2\theta}\, \del_\phi^2\,.\nn
\ee
}
\bea
\hat\del_{s}^2\equiv \hat g^{mn}\del_m\, \del_n  
&=& Y_1^2 \, (K_{13}^2 + K_{14}^2 + K_{23}^2 + K_{24}^2) +
\tY_1^2 \, (K_{57}^2 + K_{58}^2 + K_{67}^2 + K_{68}^2)\nn\\  
&&+Y_2^2 \, (K_{15}^2 + K_{16}^2 + K_{25}^2 + K_{26}^2)
+ \tY_2^2 \, (K_{37}^2 + K_{38}^2 + K_{47}^2 + K_{48}^2)\nn\\
&&+ Y_3^2 \, (K_{17}^2 + K_{18}^2 + K_{27}^2 + K_{28}^2) 
+ \tY_3^2 \, (K_{35}^2 + K_{36}^2 + K_{45}^2 + K_{46}^2)\nn\\
&&+ Y_1^2\, Y_2^2\, Y_3^2\, K_{12}^2 + 
    Y_1^2\, \tY_2^2\, \tY_3^2\, K_{34}^2 + 
    \tY_1^2\, Y_2^2\, \tY_3^2\, K_{56}^2 +
    \tY_1^2\, \tY_2^2\, Y_3^2\, K_{78}^2 \nn\\
&& -2 b_2\, b_3\, (Y_1^2\, K_{12}\, K_{34} - \tY_1^2\, K_{56}\,
K_{78})\nn\\
&& -2 b_1\, b_3\, (Y_2^2\, K_{12}\, K_{56} - \tY_2^2\, K_{34}\,
K_{78})\nn\\
&& -2 b_1\, b_2\, (Y_3^2\, K_{12}\, K_{78} - \tY_3^2\, K_{34}\,
K_{56})\,.\label{inversemet}
\eea

   In order to proceed further, it is useful to look at the geometry
of the 7-sphere in some detail.  Some useful results on this topic are
collected in Appendix B.

\subsection{The Metric Ansatz for the three dilaton/axion pairs}

    From the results in Appendix B, it follows that the inverse metric
(\ref{inversemet}) for the system with 3 dilatons and 3 axions is
a direct sum of a $4\times 4$ part involving the $\del_{\phi_i}$ basis
vectors, and a $3\times 3$ part involving the $\del_{\mu_i}$ basis
vectors (which are constrained by the fact that $\mu_i\, \mu_i=1$):
\be
\hat \del_{s}^2 = \hat\del_{s_4}^2 + \hat\del_{s_3}^2\,.
\ee
For the $4\times 4$ inverse metric, we find
\bea
\hat\del_{s_4}^2 &=& \sum_i \mu_i^{-2}\, Q_i\, \del_{\phi_i}^2
-2 b_2\, b_3\, (Y_1^2\, \del_{\phi_1}\, \del_{\phi_2} -
\tY_1^2\, \del_{\phi_3}\, \del_{\phi_4})\nn\\
&&
-2 b_1\, b_3\, (Y_2^2\, \del_{\phi_1}\, \del_{\phi_3} -
\tY_2^2\, \del_{\phi_2}\, \del_{\phi_4})
-2 b_1\, b_2\, (Y_3^2\, \del_{\phi_1}\, \del_{\phi_4} -
\tY_3^2\, \del_{\phi_2}\, \del_{\phi_3})\,,\label{inv4met}
\eea
where
\bea
Q_1 &=&Y_1^2\, Y_2^2\, Y_3^2\, \mu_1^2  +  Y_1^2\, \mu_2^2
+Y_2^2\, \mu_3^2 + Y_3^2\, \mu_4^2\,,\nn\\
Q_2 &=&Y_1^2\, \tY_2^2\, \tY_3^2\, \mu_2^2  +  Y_1^2\, \mu_1^2
+\tY_3^2\, \mu_3^2 + \tY_2^2\, \mu_4^2\,,\nn\\
Q_3 &=&Y_2^2\, \tY_1^2\, \tY_3^2\, \mu_3^2  +  Y_2^2\, \mu_1^2
+\tY_3^2\, \mu_2^2 + \tY_1^2\, \mu_4^2\,,\nn\\
Q_4 &=&Y_3^2\, \tY_1^2\, \tY_2^2\, \mu_4^2  +  Y_3^2\, \mu_1^2
+\tY_2^2\, \mu_2^2 + \tY_1^2\, \mu_3^2\,.\label{inv3met}
\eea

    For the $3\times3$ part, we find
\bea
\hat\del_{s_3}^2 &=& Y_1^2\, (\mu_1\, \del_{\mu_2} - \mu_2\,\del_{\mu_1} )^2+
 Y_2^2\, (\mu_1\, \del_{\mu_3} - \mu_3\,\del_{\mu_1} )^2
+Y_3^2\, (\mu_1\, \del_{\mu_4} - \mu_4\,\del_{\mu_1} )^2 \label{inverse6}\\
&&
 + \tY_1^2\, (\mu_3\, \del_{\mu_4} - \mu_4\,\del_{\mu_3} )^2 
 + \tY_2^2\, (\mu_2\, \del_{\mu_4} - \mu_4\,\del_{\mu_2} )^2
 + \tY_3^2\, (\mu_2\, \del_{\mu_3} - \mu_3\,\del_{\mu_2} )^2\,,\nn
\eea
where $\mu_i\, \mu_i=1$. 

   Because of the block-diagonal structure, we can invert the two
parts separately.   For the $4\times 4$ part, we straightforwardly
invert the inverse metric to obtain
\bea
d\hat s_4^2 &=& \fft1{\Xi}\,\Big[ \sum_i \mu_i^2\, Z_i\, d\phi_i^2 
+ 2 b_2\, b_3\, (\mu_1^2\, \mu_2^2\, d\phi_1\, d\phi_2
 - \mu_3^2\, \mu_4^2\, d\phi_3\, d\phi_4)\label{4x4part} \\
&&\qquad 
+ 2b_1\, b_3\, (\mu_1^2\, \mu_3^2\, d\phi_1\, d\phi_3
 - \mu_2^2\, \mu_4^2\, d\phi_2\, d\phi_4)\nn\\
&&\qquad 
+ 2 b_1\, b_2\, (\mu_1^2\, \mu_4^2\, d\phi_1\, d\phi_4
 - \mu_2^2\, \mu_3^2\, d\phi_2\, d\phi_3)\Big]\,,\nn
\eea
where
\bea
Z_1 &=& \mu_1^2 + \tY_2^2\, \tY_3^2\, \mu_2^2 + \tY_1^2\,
\tY_3^2\, \mu_3^2 + \tY_1^2\, \tY_2^2\, \mu_4^2\,,\nn\\
Z_2 &=& \mu_2^2 + Y_2^2\, Y_3^2\, \mu_1^2 + \tY_1^2\,
Y_2^2\, \mu_3^2 + \tY_1^2\, Y_3^2\, \mu_4^2\,,\nn\\
Z_3 &=& \mu_3^2 + Y_1^2\, Y_3^2\, \mu_1^2 + Y_1^2\,
\tY_2^2\, \mu_2^2 + Y_3^2\, \tY_2^2\, \mu_4^2\,,\nn\\
Z_4 &=& \mu_4^2 + Y_1^2\, Y_2^2\, \mu_1^2 + Y_1^2\,
\tY_3^2\, \mu_2^2 + Y_2^2\, \tY_3^2\, \mu_3^2\,.
\eea
The function $\Xi$ is given by
\bea
\Xi&=& Y_1^2\, Y_2^2\, Y_3^2\, \mu_1^4 + 
     Y_1^2\, \tY_2^2\, \tY_3^2\, \mu_2^4+
      \tY_1^2\, Y_2^2\, \tY_3^2\, \mu_3^4
      +\tY_1^2\, \tY_2^2\, Y_3^2\, \mu_4^4\nn\\
&& + (Y_2^2\, \tY_2^2 + Y_3^2\, \tY_3^2)\, (Y_1^2\,  \mu_1^2\, \mu_2^2
  + \tY_1^2\, \mu_3^2\, \mu_4^2)\nn\\
&& + (Y_1^2\, \tY_1^2 + Y_3^2\, \tY_3^2)\, (Y_2^2\,  \mu_1^2\, \mu_3^2
  + \tY_2^2\, \mu_2^2\, \mu_4^2)\nn\\
&& + (Y_1^2\, \tY_1^2 + Y_2^2\, \tY_2^2)\, (Y_3^2\,  \mu_1^2\, \mu_4^2
  + \tY_3^2\, \mu_2^2\, \mu_3^2)\,.
\eea

    There remains the problem of inverting the $3\times3$ part $\hat
\del_{s_3}^2$ of the inverse metric.  Since we know the
inverse metric in the form (\ref{inverse6}), expressed in
terms of the four $\del_{\mu_i}$ basis vectors formed from the  
the constrained $\mu_i$, it is helpul first to solve the constraint 
$\mu_i\, \mu_i=1$ explicitly, by introducing three angular
coordinates as follows:
\be
\mu_1=c\, \cos\ft12\theta\,,\qquad \mu_2 = c\,
\sin\ft12\theta\,,\qquad
\mu_3=s\, \cos\ft12\td\theta\,,\qquad \mu_4 = s\,
\sin\ft12\td\theta\,,\label{muxi2}
\ee
where $c=\cos\xi$, $s=\sin\xi$.  It then follows that
\bea
\del_\theta &=& \ft12 (\mu_1\, \del_{\mu_2} - \mu_2\,
\del_{\mu_1})\,,\nn\\
\del_{\td\theta} &=& \ft12 (\mu_3\, \del_{\mu_4} - \mu_4\,
\del_{\mu_3})\,,\nn\\
\del_\xi &=& -s\, c^{-1}\, (\mu_1\, \del_{\mu_1} +\mu_2\, \del_{\mu_2}) 
 + c\, s^{-1}\,  (\mu_3\, \del_{\mu_3} +\mu_4\, \del_{\mu_4}) \,.
\eea
Substituting into (\ref{inverse6}), the inverse metric is then expressed in
terms of the three unconstrained basis vectors 
$(\del_\xi, \del_\theta, \del_{\td\theta})$, and hence it can be
straightforwardly inverted.  Having done so, the downstairs metric can
then be re-expressed elegantly in terms of the redundant set of four
$d\mu_i$ differentials, in the form
\bea
d\hat s_3^2 &=& \fft1{\Xi}\, \Big[\sum_i Z_i\, d\mu_i^2  +
\ft12 b_1^2\, \Big( (\mu_1\, d\mu_1 + \mu_2\, d\mu_2)^2 +
                     (\mu_3\, d\mu_3 + \mu_4\, d\mu_4)^2\Big)\nn\\
&&
\qquad+\ft12 b_2^2\, \Big( (\mu_1\, d\mu_1 + \mu_3\, d\mu_3)^2 +
                     (\mu_2\, d\mu_2 + \mu_4\, d\mu_4)^2\Big)\nn\\
&&\qquad +\ft12 b_3^2\, \Big( (\mu_1\, d\mu_1 + \mu_4\, d\mu_4)^2 +
                     (\mu_2\, d\mu_2 + \mu_3\, d\mu_3)^2\Big)\Big]\,.
\eea

   Finally, adding this to the $4\times 4$ metric $d\hat s_4^2$ given
in (\ref{4x4part}),  we obtain the result for the downstairs 7-metric,
$d\hat s_7^2 = d\hat s_4^2 + d\hat s_3^2$:
\bea
d\hat s_7^2 &=& \fft1{\Xi}\, \Big[ \sum_i Z_i\, (d\mu_i^2 + \mu_i^2\,
d\phi_i^2) + 2b_2\, b_3\, (\mu_1^2\, \mu_2^2\, d\phi_1\, d\phi_2 
- \mu_3^2\, \mu_4^2\, d\phi_3\, d\phi_4)\nn\\
&&+ 2b_1\, b_3\, (\mu_1^2\, \mu_3^2 \,d\phi_1\, d\phi_3 
- \mu_2^2\, \mu_4^2\, d\phi_2\, d\phi_4) 
+ 2b_1\, b_2\, (\mu_1^2\, \mu_4^2\, d\phi_1\, d\phi_4 
- \mu_2^2\, \mu_3^2\, d\phi_2\, d\phi_3)\nn\\
&&+\ft12 b_1^2\, \Big( (\mu_1\, d\mu_1 + \mu_2\, d\mu_2)^2 + 
                     (\mu_3\, d\mu_3 + \mu_4\, d\mu_4)^2\Big)\nn\\
&&+\ft12 b_2^2\, \Big( (\mu_1\, d\mu_1 + \mu_3\, d\mu_3)^2 + 
                     (\mu_2\, d\mu_2 + \mu_4\, d\mu_4)^2\Big)\nn\\
&&+\ft12 b_3^2\, \Big( (\mu_1\, d\mu_1 + \mu_4\, d\mu_4)^2 + 
                     (\mu_2\, d\mu_2 + \mu_3\, d\mu_3)^2\Big)\Big]\,.
\label{d7res}
\eea

    We can now work out the eleven-dimensional metric Ansatz, given by
(\ref{d11met}).   To do this, we first note that the determinant of
(\ref{d7res}), where it is understood that the $\mu_i$ coordinates are
expressed in terms of $(\xi,\theta,\td\theta)$ using (\ref{muxi2}), is
\be
\det(\hat g_{mn}) = 
\Big(\fft{\mu_1^2\, \mu_2^2\, \mu_3^2\, \mu_4^2}{\Xi^2}\Big)\,
\Big(\fft{s^2\, c^2}{16 \Xi}\Big) = 
\fft{ \mu_1^2\, \mu_2^2\, \mu_3^2\, \mu_4^2\, s^2\, c^2}{16 \Xi^3}\,,
\ee
where in the first expression, 
the first factor is the determinant of $4\times 4$ block
involving the $\phi_i$ coordinates, and the second factor is from the
$3\times 3$ block involving the $(\xi,\theta,\td\theta)$ coordinates.  
From (\ref{delhat}), it follows that
\be
\hat\Delta= \Xi^{-\fft13}\,,
\ee
and hence from (\ref{d11met}) that the Ansatz for the
eleven-dimensional metric takes the following rather explicit form:
\bea
d\hat s_{11}^2 &=& \Xi^{\fft13}\, ds_4^2 + \Xi^{\fft13}\, d\hat s_7^2
\nn\\
&=& \Xi^{\fft13}\, ds_4^2 + g^{-2}\, \Xi^{-\fft23}\, 
\Big[ \sum_i Z_i\, (d\mu_i^2 + \mu_i^2\,
d\phi_i^2) + 2b_2\, b_3\, (\mu_1^2\, \mu_2^2\, d\phi_1\, d\phi_2 
- \mu_3^2\, \mu_4^2\, d\phi_3\, d\phi_4)\nn\\
&&+ 2b_1\, b_3\, (\mu_1^2\, \mu_3^2 \,d\phi_1\, d\phi_3 
- \mu_2^2\, \mu_4^2\, d\phi_2\, d\phi_4) 
+ 2b_1\, b_2\, (\mu_1^2\, \mu_4^2\, d\phi_1\, d\phi_4 
- \mu_2^2\, \mu_3^2\, d\phi_2\, d\phi_3)\nn\\
&&+\ft12 b_1^2\, \Big( (\mu_1\, d\mu_1 + \mu_2\, d\mu_2)^2 + 
                     (\mu_3\, d\mu_3 + \mu_4\, d\mu_4)^2\Big)\nn\\
&&+\ft12 b_2^2\, \Big( (\mu_1\, d\mu_1 + \mu_3\, d\mu_3)^2 + 
                     (\mu_2\, d\mu_2 + \mu_4\, d\mu_4)^2\Big)\nn\\
&&+\ft12 b_3^2\, \Big( (\mu_1\, d\mu_1 + \mu_4\, d\mu_4)^2 + 
                     (\mu_2\, d\mu_2 + \mu_3\, d\mu_3)^2\Big)\Big]\,.
\label{d11metricans}
\eea
Note that we have reinstated the gauge-coupling constant $g$ in this
expression. 

    Having obtained the Kaluza-Klein metric Ansatz for the three
dilaton/axion pairs, it is a simple matter to incorporate also the
associated $U(1)^4$ gauge fields that naturally accompany this
truncation of the maximal supergravity.  Denoting their potentials by
$A_\1^i$, for $i=1,2,3,4$, we simply replace each occurrence of
$d\phi_i$ in (\ref{d11metricans}) by
\be
d\phi_i \longrightarrow d\phi_i - g\,  A_\1^i\,.
\ee

   Finally in this section, we may note that our result (\ref{d11metricans})
is consistent with previously-obtained special cases.  In particular,
if we set the three axions $\chi_i$ to zero, then the function $\Xi$ reduces
to
\be
\Xi= \Delta^2\,,
\ee
where 
\be
\Delta = Y_1\, Y_2\, Y_3 \, \mu_1^2 + \fft{Y_1}{Y_2\, Y_3}\, \mu_2^2 +
 \fft{Y_2}{Y_1\, Y_3}\, \mu_3^2 +  \fft{Y_3}{Y_1\, Y_2}\, \mu_4^2 \,.
\ee
In the absence of axions, it is natural to define
\be
X_1=  Y_1\, Y_2\, Y_3\,,\quad X_2 = \fft{Y_1}{Y_2\, Y_3}\,,\quad
X_3 = \fft{Y_2}{Y_1\, Y_3}\,,\quad X_4=\fft{Y_3}{Y_1\, Y_2}\,,
\ee
implying that we shall have
\be
\Delta = \sum_i X_i\, \mu_i^2\,,\qquad Z_i= \Delta\, X_i^{-1}\,.
\ee
It can be seen that the metric Ansatz (\ref{d11metricans}) therefore indeed
reduces to the one given in \cite{ten} if the axions are set to zero.

\subsection{The Ansatz for the 4-form Field Strength}

    In principle, we should like to obtain also the Ansatz for the
4-form field strength $\hat F_\4$ of eleven-dimensional supergravity.
In spherical Kaluza-Klein reductions it is always much more difficult
to obtain the Ansatz for antisymmetric tensors than for the metric,
and the present case is no exception.  Unfortunately, one can only
obtain limited guidance from those results that are presented in
\cite{dwn}.  In other truncations, simpler than the case in hand, it
has been possible to determine the field-strength Ansatz by
brute-force methods, and up to a point, this technique is still useful
here.  (This method was used successfully in \cite{d4gauge}, where the
complete and explicit Ans\"atze for the $S^7$ reduction to the bosonic
sector of $N=4$, $SO(4)$ gauged supergravity in $D=4$ were obtained.)
The contributions to the 4-form Ansatz can be organised into different
sectors, and in all except one of these we have obtained complete
results.  Since these are instructive and useful in their own right,
it seems to be worthwhile to present those results that we have
obtained here.

   We begin with a summary of the four-dimensional theory comprising
gravity, the three dilaton/axion pairs, and the associated $U(1)^4$
gauge fields.

\subsubsection{$D=4$ Lagrangian}

    The complete Lagrangian for four-dimensional $N=8$ $SO(8)$-gauged
supergravity was obtained in \cite{dwn00,dwn0}.  In \cite{duliu,ten},
the truncation to the $N=2$ $U(1)^4$-gauged subsector was discussed.
Adapting these results to the notation of this paper, we find that
the four-dimensional bosonic Lagrangian for this $N=2$ truncation is given by
\be
{\cal L}_4 = R\, {*\oneone} - \ft12 \sum_{i=1}^3(
{*d\varphi_i}\wedge d\varphi_i + e^{2\varphi_i}\, {*d\chi_i}\wedge
d\chi_i\Big) -V\, {*\oneone} + {\cal L}_{Kin} + {\cal L}_{CS}\,,\label{d4lag}
\ee
where $V$ is the potential for the scalar fields, and 
${\cal L}_{Kin}$ and ${\cal L}_{CS}$ are the kinetic terms and
the Chern-Simons terms for the four $U(1)$ gauge fields
$F_\2^i=dA_\1^i$.  The scalar potential is given by
\be
V  = - 4 g^2\, \sum_{i=1}^3 (Y_i^2 + \tY_i^2)\,.\label{scalpot}
\ee

   The kinetic terms for the gauge fields are 
\bea
{\cal L}_{Kin} &=& 
-\ft12|W|^{-2}\, \Big[P_0\,\Big(\tY_1^2\, \tY_2^2\, \tY_3^2\,
{*F_\2^1}\wedge F_\2^1 +
\tY_1^2\, Y_2^2\, Y_3^2\, {*F_\2^2}\wedge F_\2^2 \nn\\
&&\qquad\qquad\quad + Y_1^2\, \tY_2^2\, Y_3^2\, {*F_\2^3}\wedge F_\2^3 
+ Y_1^2\, Y_2^2\, \tY_3^2\, {*F_\2^4}\wedge F_\2^4\Big)\nn\\
&& \qquad\qquad+ 2 P_1\, b_2\, b_3\, (\tY_1^2\, {*F_\2^1}\wedge F_\2^2 
-Y_1^2\, {*F_\2^3}\wedge F_\2^4)\nn\\
&& \qquad\qquad + 2 P_2\, b_1\, b_3\, (\tY_2^2\, {*F_\2^1}\wedge F_\2^3 
-Y_2^2\, {*F_\2^2}\wedge F_\2^4)\nn\\
&& \qquad\qquad + 2 P_3\, b_1\, b_2\, (\tY_3^2\, {*F_\2^1}\wedge F_\2^4 
-Y_3^2\, {*F_\2^2}\wedge F_\2^3)\Big]\,,\label{lkin}
\eea
where
\bea
&&P_0 \equiv 1 +b_1^2 + b_2^2 +b_3^2\,,\quad
W \equiv P_0 - 2\im\, b_1\, b_2\, b_3\,,\nn\\
&&
P_1\equiv  1 -b_1^2 + b_2^2 +b_3^2\,,\quad
P_2\equiv  1 +b_1^2 - b_2^2 +b_3^2\,,\quad
P_3\equiv  1 +b_1^2 + b_2^2 -b_3^2\,.
\eea

    Finally, the Chern-Simons terms for the gauge fields are
\bea
{\cal L}_{CS} &=& -|W|^{-2}\, \Big[ b_1\, b_2\, b_3\, 
\Big(\tY_1^2\, \tY_2^2\, \tY_3^2\, F_\2^1\wedge F_\2^1 +
\tY_1^2\, Y_2^2\, Y_3^2\, F_\2^2\wedge F_\2^2\nn\\
&& \qquad\qquad\qquad\quad + Y_1^2\, \tY_2^2\, Y_3^2\, F_\2^3\wedge F_\2^3 
+ Y_1^2\, Y_2^2\, \tY_3^2\, F_\2^4\wedge F_\2^4\Big)\nn\\
&&
\qquad\qquad
+b_1\, ( P_0 + 2 b_2^2\, b_3^2)\, (\tY_1^2\, F_\2^1\wedge F_\2^2 
-Y_1^2\, F_\2^3\wedge F_\2^4)\nn\\
&&\qquad\qquad
+ b_2\, (P_0 + 2 b_1^2\, b_3^2)\, (\tY_2^2\, F_\2^1\wedge F_\2^3 
-Y_2^2\, F_\2^2\wedge F_\2^4)\nn\\
&&
\qquad\qquad
+b_3\, (P_0 + 2 b_1^2\, b_2^2)\, (\tY_3^2\, F_\2^1\wedge F_\2^4 
-Y_3^2\, F_\2^2\wedge F_\2^3)\Big]\,.\label{lcs}
\eea

    From (\ref{d4lag}), we find that the equations of motion for the
gauge fields are
\be
d(|W|^{-2}\, R_i)=0\,,\label{gaugeeom}
\ee
for $i=1,2,3,4$, where
\bea
R_1 &=& 
\tY_1^2\, \tY_2^2\, \tY_3^2\, [P_0\, {*F_\2^1} + 2 b_1\, b_2\, b_3\, F_\2^1] 
+\tY_1^2\, 
[P_1\, b_2\, b_3\, {*F_\2^2} + b_1\,(P_0 + 2 b_2^2\, b_3^2)\, F_\2^2]
\label{ridef}\\
&&
+\tY_2^2 \,
[P_2\, b_1\, b_3\, {*F_\2^3} + b_2\,(P_0 + 2 b_1^2\, b_3^2)\, F_\2^3]
+\tY_3^2 \,
[P_3\, b_1\, b_2\, {*F_\2^4} + b_3\,(P_0 + 2 b_1^2\, b_2^2)\, F_\2^4]
\,,\nn\\
R_2 &=&
\tY_1^2\, Y_2^2\, Y_3^2\, [P_0\, {*F_\2^2} + 2 b_1\, b_2\, b_3\, F_\2^2] 
+\tY_1^2\, 
[P_1\, b_2\, b_3\, {*F_\2^1} + b_1\,(P_0 + 2 b_2^2\, b_3^2)\, F_\2^1]
\nn\\
&&
-Y_2^2 \,
[P_2\, b_1\, b_3\, {*F_\2^4} + b_2\,(P_0 + 2 b_1^2\, b_3^2)\, F_\2^4]
-Y_3^2 \,
[P_3\, b_1\, b_2\, {*F_\2^3} + b_3\,(P_0 + 2 b_1^2\, b_2^2)\, F_\2^3]
\,,\nn\\
R_3 &=&
Y_1^2\, \tY_2^2\, Y_3^2\, [P_0\, {*F_\2^3} + 2 b_1\, b_2\, b_3\, F_\2^3] 
-Y_1^2\, 
[P_1\, b_2\, b_3\, {*F_\2^4} + b_1\,(P_0 + 2 b_2^2\, b_3^2)\, F_\2^4]
\nn\\
&&
+\tY_2^2 \,
[P_2\, b_1\, b_3\, {*F_\2^1} + b_2\,(P_0 + 2 b_1^2\, b_3^2)\, F_\2^1]
-Y_3^2 \,
[P_3\, b_1\, b_2\, {*F_\2^2} + b_3\,(P_0 + 2 b_1^2\, b_2^2)\, F_\2^2]
\,,\nn\\
R_4 &=&
Y_1^2\, Y_2^2\, \tY_3^2\, [P_0\, {*F_\2^4} + 2 b_1\, b_2\, b_3\, F_\2^4] 
-Y_1^2 \,
[P_1\, b_2\, b_3\, {*F_\2^3} + b_1\,(P_0 + 2 b_2^2\, b_3^2)\, F_\2^3]
\nn\\
&&
-Y_2^2 \,
[P_2\, b_1\, b_3\, {*F_\2^2} + b_2\,(P_0 + 2 b_1^2\, b_3^2)\, F_\2^2]
+\tY_3^2\,
[P_3\, b_1\, b_2\, {*F_\2^1} + b_3\,(P_0 + 2 b_1^2\, b_2^2)\, F_\2^1]
\,.\nn
\eea

\subsubsection{The Ansatz for $\hat F_\4$}

    In previous papers the Ansatz for the 4-form field strength $\hat
F_\4$ was obtained for the $U(1)^4$ truncation in absence of the three
axions \cite{ten}, and for the $N=4$ gauged $SO(4)$ truncation, in
which there is one scalar and one axion \cite{d4gauge}.  Based on
those results, it can be seen to be natural to write the Ansatz for
$\hat F_\4$ as the sum of three terms, each with its own
characteristic contribution to the whole. 

   Thus we are led to the following construction for the 4-form field
strength:
\bea
\hat F_\4 &=& -2 g\, U\, \ep_\4 + \hat F_\4' + \hat F_\4''
\nn\\
&&+\ft1{2 g}\,(2 Y_1^{-1}\,
{*dY_1} -\chi_1\, Y_1^4\, {*d\chi_1})\wedge d(\mu_1^2 + \mu_2^2) \nn\\
&&+\ft1{2\, g}\,(2 Y_2^{-1}\,
{*dY_2} -\chi_2\, Y_2^4\, {*d\chi_2})\wedge d(\mu_1^2 + \mu_3^2) \nn\\
&&+\ft1{2 g}\,(2 Y_3^{-1}\,
{*dY_3} -\chi_3\, Y_3^4\, {*d\chi_3})\wedge d(\mu_1^2 + \mu_4^2)
\,,\label{f4ans1}
\eea
where
\be
U = Y_1^2\, (\mu_1^2 + \mu_2^2) + \tY_1^2\,  (\mu_3^2 +
\mu_4^2)
 +Y_2^2\, (\mu_1^2 + \mu_3^2) + \tY_2^2\,  (\mu_2^2 +
\mu_4^2)
 +Y_3^2\, (\mu_1^2 + \mu_4^2) + \tY_3^2\,  (\mu_2^2 +
\mu_3^2)\,,\label{udef}
\ee
and $\ep_\4$ denotes the volume form on the four-dimensional spacetime.
The term $\hat F_\4''$ will be given by
\be
\hat F_\4'' = - \fft1{2 g^2}\, |W|^{-2} \, \sum_i d\mu_i^2\wedge 
(d\phi_i - g\, A_\1^i)\wedge R_i\,.\label{fppans}
\ee
(We shall justify these expressions below.)
The remaining term is $\hat F_\4'$.  This will be written in terms of
a potential $\hat A_\3'$, as $\hat F_\4' = d\hat A_\3'$.  It will be
the determination of $\hat A_\3'$ that presents the greatest
difficulty.

   It will be noted that $\hat F_\4$ does not identically satisfy
$d\hat F_\4=0$.  This feature was already seen in the truncations in
\cite{ten} and \cite{d4gauge}.  It is not possible, at least within
the usual second-order formulation of eleven-dimensional
supergravity, to write an Ansatz for $\hat F_\4$ in the $S^7$ reduction
that identically satisfies $d\hat F_\4=0$.  An implication from this
is that one cannot write the Ansatz directly on the potential $\hat
A_\3$, which in turn means that one cannot write an Ansatz that can be
substituted directly into the eleven-dimensional action.  One must
work at the level of the equations of motion.

   In fact the requirement that $\hat F_\4$ must satisfy the Bianchi
identity $d\hat F_\4=0$ provides us with very important clues as to
the correct form of the reduction Ansatz, and we used this in writing
down our results in (\ref{f4ans1}) and (\ref{fppans}).  The point is
that the Bianchi identity will be satisfied by virtue of the $D=4$
equations of motion for the scalar fields and the $U(1)$ gauge fields
being satisfied.  (To be precise, the scalar equations of motion in
question here are those of the three dilatons $\varphi_i$, in
combination with certain non-linear admixtures of the three axion
equations of motion.)  Of course the contribution to $\hat F_\4$ from
$\hat A_\3'$, whose precise form we have not been able to determine,
does not enter into the discussion of the Bianchi identity, since it
gives a contribution $\hat F_\4'$ that identically satisfies $d\hat
F_\4'=0$. 

   To see how the Bianchi identity $d\hat F_\4=0$ implies the
four-dimensional equations of motion for the scalars and the gauge
fields, we note from the structure of (\ref{f4ans1}) and (\ref{fppans})
that after acting with $d$ we shall have two distinct classes of
terms.  First, there will be terms of the form $d\mu_i^2\wedge
\omega_\4$, where $\omega_\4$ is a 4-form living entirely in the
four-dimensional spacetime.  ($\omega_\4$ will comprise terms of the
form $\ep_\4$, and of the form $d{*d Y_i}$, \etc  Of course they are
all proportional to $\ep_\4$.)  The requirement of the 
vanishing of these terms will imply the scalar equations of motion.
Secondly, there will be terms of the form $d\mu_i^2 \wedge
(d\phi_i-\ft12 g\, A_\1^i)\wedge \omega_\3$ coming from the action of
$d$ on $\hat F_\4''$, where $\omega_\3$ is a 3-form living in the
four-dimensional spacetime.  The vanishing of these terms will imply
the four-dimensional equations of motion for the gauge fields.

    Let us consider the second type of contribution first, since it is
the simpler one.  The terms of this type come only from $d\hat
F_\4''$, and give
\be
\sum_i d\mu_i^2\wedge (d\phi_i- g\, A_\1^i)\wedge
d(|W|^{-2}\, R_i)=0\,.
\ee
This can immediately be seen to imply precisely the equations of
motion for the four $U(1)$ gauge fields, given in (\ref{gaugeeom}).

   It remains to check that the terms of the form $d\mu_i^2\wedge
\omega_\4$ coming from the Bianchi identity vanish by virtue of the
four-dimensional scalar equations of motion.  The kinetic terms of
these scalar equations come from the action of $d$ on the final three
lines in (\ref{f4ans1}).  Clearly, we get the combinations of the form
\be
d( 2Y_1^{-1}\, {*dY_1} - \chi_1\, Y_1^4\, {*d\chi_1})\,,
\ee
arising (with similar independent expressions involving the
$(Y_2,\chi_2)$ and $(Y_3,\chi_3)$ pairs).  This is a combination of the
$\varphi_1$ and the $\chi_1$ equations of motion.  In fact it is
\be
[d{*d\varphi_1} + e^{2\varphi_1}\, {*d\chi_1}\wedge d\chi_1]  -
\chi_1\, [d(e^{2\varphi_1}\, {*d\chi_1})]\,,
\ee
where the first quantity in square brackets is the dilaton equation of
motion, and the second quantity in square brackets is the axion
equation of motion. 

    This particular combination, of the dilaton equation plus an
admixture of the axion equation, is an especially simple one to
compare with the scalar equations of motion coming from the
four-dimensional Lagrangian (\ref{d4lag}).  It means that we are
looking at the combination that comes from the following variation of
the $D=4$ Lagrangian:
\be
\hat\delta {\cal L}_4 \equiv \fft{\delta{\cal L}_4}{\delta \varphi_1} -
\chi_1\, \fft{{\cal L}_4}{\delta\chi_1}\,.
\ee
If we define a symbol $\hat\delta$ to denote this specific
combination of field variations, \ie
\be
\hat\delta  \equiv \fft{\delta}{\delta \varphi_1} -
\chi_1\, \fft{\delta}{\delta\chi_1}\,,
\ee
then we find the great simplification that
\be
\hat\delta Y_1^2 = Y_1^2\,,\qquad \hat\delta \tY_1^2 = -\tY_1^2\,,
\qquad \hat\delta b_1 = 0\,.\label{hatvar}
\ee
(Of course since we are focusing on the scalars with the index $i=1$
at the moment, all of the scalar quantities with $i=2$ or $i=3$ labels
are invariant under this transformation.)  The last equation in
(\ref{hatvar}), $\hat\delta b_1=0$, leads to an {\it enormous}
simplification when we vary ${\cal L}_{Kin}$ and ${\cal L}_{CS}$ given
by (\ref{lkin}) and (\ref{lcs}).  It means that $|W|$, the
$P_a$, and all the $b_i$ are invariant.  We need
only consider $Y_1$ and $\tY_1$, and these just vary by the
very simple rules given in (\ref{hatvar}).

    With these observations, it becomes a relatively straightforward
matter to verify that the terms of the form $d\mu_i^2\wedge \omega_\4$
that arise in the Bianchi identity for $\hat F_\4$ vanish {\it
precisely} as a consequence of the scalar equations of motion
following from (\ref{d4lag}), to all orders in scalar fields and gauge
field strengths.  Note that the contributions to the scalar equations
of motion from the potential $V$ given in (\ref{scalpot}) arise from
the action of the exterior derivative on the term $-2g \,U\,
\ep_\4$ in (\ref{f4ans1}).  This part of the calculation can be seen
quite easily, and can be examined in isolation from the more
complicated contributions from the four-dimensional gauge fields.

    The contribution $\hat F_\4'=\hat A_\3'$ in
(\ref{f4ans1}) remains undetermined.  We know some aspects of it
structure, for example that it is of the general from
\be
\hat A_\3' = \sum_{i\ne j} h_{ij}\, (\mu_i^2\, d\mu_j^2 -\mu_j^2\,
d\mu_i^2)\wedge (d\phi_i - g\, A_\1^i)\wedge  
(d\phi_j - g\, A_\1^j)\,,\label{aprime}
\ee
where the functions $h_{ij}$ depend on the scalars $\varphi_i$ and
$\chi_i$, and the direction cosines $\mu_i$.  At leading order, these
terms will give rise to the linearised Ansatz for the axions
$\chi_i$.  If explicit expressions for the complete Ansatz for the
$N=8$ $SO(8)$ gauged supergravity embedding were available, $A_\3'$
could in principle be determined by substituting the expressions for
$u_{ij}{}^{KL}$ and $v_{ijKL}$ appearing in (\ref{uexpr}) and
(\ref{vexpr}) into them.  To the extent that such expressions are
implicit in the work of \cite{dwn}, a procedure in principle exists
for reading off $A_\3'$.  It is not clear that attempting such a
substitution would be simpler than a brute-force direct attack on the
problem, of the type that has proved successful in previous (simpler)
cases \cite{ten,d4gauge}.

\subsection{Domain wall solutions and their oxidation}

      The four-dimensional $U(1)^4$ Lagrangian (\ref{d4lag}) supports a
four-charge AdS black hole solution \cite{duliu}.  In the extremal limit,
the four $U(1)$ gauge fields decouple and the solution becomes
AdS domain wall, supported by the scalar fields only.  
It is given by \cite{dist}
\bea
ds_{4}&=& (gr)^4 (H_1H_2H_3H_4)^{-1/2}\, dx^\mu\, dx_\mu +
(H_1 H_2 H_3 H_4)^{1/2}\, \fft{dr^2}{g^2 r^2}\,,\nn\\
e^{\varphi_i} &=& Y_1^2 = f_i\,,\qquad
\chi_i=0\,,
\eea
where
\bea
f_1&=& \fft{(H_3 H_4)^{1/2}}{(H_1 H_2)^{1/2}}\,,\qquad
f_2= \fft{(H_2 H_4)^{1/2}}{(H_1 H_3)^{1/2}}\,,\nn\\
f_3&=& \fft{(H_2 H_3)^{1/2}}{(H_1 H_4)^{1/2}}\,,\qquad
H_i=1 + \fft{\ell_i^2}{r^2}\,.
\eea
This solution can be oxidised back to $D=11$ \cite{dist}, where it
acquires the interpretation of being a continuous ellipsoidal
distribution \cite{KLT,FGPW,BS,BSI,dist,BBS} of M2-branes.

     The scalar kinetic terms in the Lagrangian (\ref{d4lag}) are
invariant under global $SL(2,R)^3$ transformations, corresponding to
the usual fractional-linear group action on each of the axion/dilaton
pairs.  The scalar potential in (\ref{d4lag}), on the other hand, is
invariant only under the $SO(2)^3$ subgroup transformations
\be
\tau_i \rightarrow \tau_i' = \fft{\cos\lambda_i\, \tau + sin\lambda_i}{
-\sin\lambda_i\, \tau + \cos\lambda_i}\,.
\ee
where $\tau_i\equiv = \chi_i + {\im }\, e^{-\varphi_i}$.  Applying
these global transformations to the original domain walls we obtain
new solutions, with 
\be
Y_i^2=e^{\varphi_i} = \fft1{f_i} 
(f_i^2 \cos^2\lambda_i + \sin^2\lambda_i)\,,\qquad
\chi_i=\fft{\ft12(f_i-1)}{f_i^2 \cos^2\lambda_i + \sin^2\lambda_i}
\,.
\ee
The $\wtd Y_i$ are hence given by
\be
\wtd Y_i^2 = \fft{f^2 + \ft14(f-1)^2 \sin^2(2\lambda_i)}{
f_i(f_i^2 \cos^2\lambda_i + \sin^2\lambda_i)}\,.
\ee

        Having obtained the $SO(2)^3$ rotated domain-wall solutions,
they can be oxidised back to $D=11$.  The eleven-dimensional metric is
given by substituting the solution into (\ref{d11metricans}).  These
solutions with non-vanishing $\chi_i$ no longer simply describe
distributed M2-branes.  To see this we note from (\ref{aprime}) that
with non-vanishing axions the field strength $F_\4$ will involve
components lying purely in the internal $S^7$.  By contrast, in a
distributed M2-brane solution one has $F_\4=d^3x\wedge dH^{-1}$, where
$H$ is the harmonic function in the transverse space.  Thus for a
distributed M2-brane the field strength $F_\4$ always carries three
world-volume indices.

\section{The 2-scalar $D=5$ embedding in type IIB}

    In this section, we consider the embedding of the 2-scalar
truncation of $D=5$ gauged supergravity discussed in the introduction,
and its embedding in the type IIB theory {\it via} an $S^5$ reduction.
In the early stages of the derivation, we retain all four of the
scalar fields of the truncation discussed in \cite{kpw}.

\subsection{The metric reduction Ansatz}

    The set of 42 spin-0 fields in the complete $SO(6)$ gauged $N=8$
supergravity in $D=5$ \cite{grw} are described by a 27-bein ${\cal
V}$, which transforms under local $USp(8)$ and global $E_6$.  The
truncation to four spin-0 fields is described in \cite{kpw}, in terms
of an $SL(6,R)\times SL(2,R)$ basis, for which the components of the
vielbein are decomposed as $({\cal V}^{IJ ab}, {\cal
V}_{I\a}{}^{ab})$.  In terms of this decomposition, the following
conjecture for the inverse $S^5$ metric has been proposed \cite{kpw}:
\be
\hat g^{mn}(x,y)\equiv \hat\Delta^{-\fft23}\, g^{mn}(x,y) 
= 2 K^m_{IJ}\, K^n_{KL}\, 
\wtd{\cal V}_{IJ ab}\, \wtd{\cal V}_{KL cd}\, \Omega^{ac}\,
\Omega^{bd}\,,\label{d5metans}
\ee
where $\wtd{\cal V}$ is the inverse of the vielbein ${\cal V}$, 
$\hat\Delta^2 = \det(g_{mn}(x,y))/\det(g_{mn}(y))$, and
$g_{mn}(y)$ is the undeformed round $S^5$ metric where the
scalar fields are set to zero.  The ten-dimensional metric Ansatz will
then be
\be
d\hat s_{10}^2 = \hat\Delta^{-\fft23}\, ds_5^2 + g_{mn}(x,y)\, dy^m\,
dy^n   =\hat\Delta^{-\fft23}\, (ds_5^2 + \hat g_{mn}(x,y)\, dy^m\,
dy^n)\,.\label{d10metans}
\ee

    The process of making the 4-scalar truncation in the vielbein
${\cal V}$ has been described in detail in \cite{fgpw}.  Substituting this
into the metric Ansatz (\ref{d5metans}) is a mechanical exercise that
is most conveniently implemented by computer.  Since the final result
is considerably simpler than the intermediate stages we shall, without
further ado, present the final answer.  We find that the inverse
5-sphere metric $\hat \del_{s_5}^2\equiv \hat g^{mn}\, \del_m\, \del_n$ is
given by
\bea
\hat\del_{s_5}^2 &=& X^{-1}\, \Big( \cosh 2 y_2\, (\cosh 2r
-\sin\theta\, \sinh 2r)\, (K_{15}^2 + K_{25}^2 + K_{35}^2 + K_{45}^2)
\nn\\
&&\qquad +\cosh 2y_2\, (\cosh 2r
+\sin\theta\, \sinh 2r)\, (K_{16}^2 + K_{26}^2 + K_{36}^2 +
K_{46}^2)\nn\\
&&\qquad + 2\cos\theta\, \sinh 2r\, \sinh 2y_2\, (K_{26}\, K_{35} -
K_{25}\, K_{36} + K_{16}\, K_{45} - K_{15}\, K_{46})\Big)\nn\\
&&+ X^2\, \Big( \ft14 (3-\cos\theta + 2\cos^2\theta\, \cosh 4r)\, 
(K_{12}^2 + K_{34}^2) + (K_{14}^2 + K_{23}^2) \nn\\
&&\qquad 
+ \cosh^2 2y_2\, (K_{13}^2 + K_{24}^2) + 2\cos^2\theta\, \sinh^2
2r\, K_{12}\, K_{34} - 2\sinh^2 2y_2\, K_{13}\, K_{24}\Big)\nn\\
&& + X^{-4}\, K_{56}^2\,.\label{d5upper}
\eea
The scalars $(X, r, y_2, \theta)$ are related to the quantities 
$(\rho, \vp_1, \vp_2, \phi)$ appearing in \cite{fgpw} by
\be
\rho= X^{-\fft12}\,,\quad
r = \ft12(\vp_2-\vp_1)\,,\quad y_2 = \ft12 (\vp_1 +\vp_2)\,,\quad 
\theta = 2\phi\,.
\ee
Note that the $D=5$ scalar Lagrangian for this truncation is
\be
{\cal L} = - 2 \sum_{i=1}^3 (\del\vp_i)^2 - \sinh^2(\vp_1-\vp_2)\,
(\del\theta)^2  - V\,,\label{d5lag}
\ee
where $X=e^{-\sqrt6\, \vp_3/2}$, and 
the scalar potential $V$ takes the form \cite{kpw}
\bea
V &=& g^2\, \Big( X^2\, [1-\cos^2\theta\, (\sinh^2\vp_1 - \sinh^2
\vp_2) ] + X^{-1}\, [\cosh 2\vp_1 +\cosh 2\vp_2] \label{d5pot}\\
&&\quad + \ft1{16}\, X^{-4}\, [2 + 2 \sin^2\theta -2\sin^2\theta\,
\cosh(2(\vp_1 -\vp_2)) -\cosh 4\vp_1 -\cosh 4\vp_2]\Big)\,.\nn
\eea

    At this stage, we impose the further truncation to the 2-scalar
subsector that we really want to consider.  This corresponds to
setting $\theta=0$ and $\vp_2=0$ \cite{fgpw}.  It is easily verified
from (\ref{d5lag}) and (\ref{d5pot}) that this is a consistent
truncation.   Thus we shall have $r=-\ft12\vp$, and $y_2=\ft12\vp$,
where we now drop the ``1'' subscript on $\vp_1$.  The potential
(\ref{d5pot}) reduces to
\be
V= \cosh^2\vp\, \Big[ X^2\, (2-\cosh^2\vp) + 2 X^{-1} -\ft12 X^{-4}\,
\sinh^2\vp\Big]\,.\label{d5pot2}
\ee
It is convenient also at this stage to perform a labelling of indices
on the Killing vectors $K_{ij}$ in (\ref{d5upper}), under which the
index values $(2,3,4)$ are cycled: $2\rightarrow 3$,  $3\rightarrow 4$
and $4\rightarrow 2$.  

    We now adopt a description of the round 5-sphere that is precisely
analogous to the one that we introduced in Appendix B for $S^7$.  This
time, we shall end up with three ``direction cosines'' $\mu_i$,
subject to the condition $\mu_i\, \mu_i=1$, and three azimuthal angles 
$\phi_i$.  After manipulations similar to those in section 2, we
arrive at the following expression for the inverse 5-sphere metric
$\hat\del_5^2$:
\be
\hat \del_{s_5}^2 = \hat \del_{s_2}^2 + \hat \del_{s_3}^2\,,
\ee
where the $2\times 2$ and $3\times3$ blocks are given by
\bea
\hat\del_{s_2}^2 &=& \cosh^2\vp\, \Big( X^{-1}\, 
[(\mu_1\, \del_{\mu_3} -\mu_3\, \del_{\mu_1})^2  +
(\mu_2\, \del_{\mu_3} -\mu_3\, \del_{\mu_2})^2] +
X^2\,  (\mu_1\, \del_{\mu_2} -\mu_2\, \del_{\mu_1})^2
\Big)\,,\nn\\
\hat\del_{s_3}^2 &=& \Delta\, \cosh^2\vp\, \Big[X\, (\mu_1^{-2}\,
\del_{\phi_1}^2 +\mu_2^{-2}\,\del_{\phi_2}^2) + X^{-2}\, \mu_3^{-2}\,
\del_{\phi_3}^2 \Big]\nn\\
&& -\sinh^2\vp\, \Big( X\, (\del_{\phi_1} + \del_{\phi_2})- X^{-2}\,
\del_{\phi_3}\Big)^2 \,, 
\eea
and
\be
\Delta \equiv (\mu_1^2+\mu_2^2)\, X + \mu_3^2\, X^{-2}\,.
\ee
Note that the $2\times2$ inverse metric $\hat\del_{s_2}^2$ is just equal
to the metric for the single-scalar truncation when $\vp=0$,
multiplied by a factor of $\cosh^2\vp$.  The $3\times 3$ inverse
metric is equal to $\cosh^2\vp$ times the $\vp=0$ metric, with the
correction term appearing in its second line.

    The inverse of the $3\times3$ block $\hat\del_{s_3}^2$ is
straightforward to calculate, and we find
\be
d\hat s_3^2 = \fft{\sech^2\vp}{\Delta}\, \Big( X^{-1}\, (\mu_1^2\,
d\phi_1^2 + \mu_2^2\, d\phi_2^2) + X^2\, \mu_3^2\, d\phi_3^2 \Big)
+ \fft{\tanh^2\vp}{\Delta^2}\, ( \mu_1^2 \, d\phi_1 
+ \mu_2^2 \, d\phi_2 - \mu_3^2 \, d\phi_3)^2\,.\\
\ee
Note that the determinant of $d\hat s_3^2$ is given by $(\mu_1\,
\mu_2\, \mu_3)^2/(\Delta^3\, \cosh^4\vp)$.

    For the $2\times2$ block, the inversion gives the metric
\be
d\hat s_2^2 = \fft{\sech^2\vp}{\Delta}\, \Big( X^{-1}\, (d\mu_1^2 +
d\mu_2^2) + X^2\, d\mu_3^2\Big)\,.
\ee

    It is helpfull at this stage to reparameterise the direction
cosines $\mu_i$, and make redefinitions of the azimuthal angles
$(\phi_1,\phi_2)$ as follows:
\bea
&&\mu_1 = \cos\xi\, \cos\ft12\vartheta\,,\qquad
\mu_2 = \cos\xi\, \sin\ft12\vartheta\,,\qquad
\mu_3= \sin\xi\,,\nn\\
&&\phi_1 = \ft12(\psi + \phi)\,,\qquad \phi_2 = \ft12(\psi -\phi)\,.
\eea
In fact $(\vartheta,\phi,\psi)$ are just the Euler angles on $S^3$.
One can define left-invariant 1-forms $\sigma_i$, as
\be
\sigma_1 + \im\, \sigma_2 = e^{-\im\psi}\, (d\vartheta + \im\,
\sin\vartheta\, d\phi)\,,\qquad \sigma_3 = d\psi + \cos\vartheta\,
d\phi\,.
\ee
These satisfy $d\sigma_1 = -\sigma_2\wedge \sigma_3$, and cyclically.
Defining also
\be
c \equiv \cos\xi\,,\qquad s\equiv \sin\xi\,,
\ee
we find that the 5-dimensional internal metric $d\hat s_5^2 \equiv
\hat g_{mn}(x,y)\, dy^m\, dy^n=d\hat
s_2^2 + d\hat s_3^2$ becomes
\bea
d\hat s_5^2 &=& \fft{\sech^2\vp}{\Delta}\, \Big[ X\, \Delta\, d\xi^2
+ \ft14 X^{-1}\, c^2\, (\sigma_1^2 + \sigma_2^2 + \sigma_3^2) + X^2\,
s^2\, d\phi_3^2\Big]\nn\\
&& + \fft{\tanh^2\vp}{4 \Delta^2}\, (c^2\, \sigma_3 - 2s^2\,
d\phi_3)^2\,,\label{d5metres1}
\eea
where
\be
\Delta = X\, c^2 + X^{-2}\, s^2\,.
\ee
In the absence of the pseudoscalar field $\vp$, this reduces to the
metric Ansatz encountered in the $N=4$ gauged $SU(2)\times U(1)$
supergravity embedding obtained in \cite{d5gauge}.  In that case, the
scalar field $X$ parameterises inhomogeneous deformations of $S^5$
viewed as a foliation of $S^3\times S^1$ surfaces. 

   With the pseudoscalar $\vp$ non-vanishing, it is advantageous to
rewrite the metric (\ref{d5metres1}) as the sum of squares of just
five quantities, by completing the square.  After doing this, we
obtain the result
\be
d\hat s_5^2 = \fft{X}{\cosh^2\vp}\, d\xi^2 + \fft{c^2\, X^{-1}}{4\Delta\,
\cosh^2\vp}\, (\sigma_1^2+\sigma_2^2) 
+ \fft{c^2\, X}{4\Omega}\, \sigma_3^2
+ \fft{s^2\, \Omega}{\Delta^2\, \cosh^2\vp}\, 
\Big(d\phi_3 - \fft{c^2\, \sinh^2\vp}{2\Omega}\, \sigma_3\Big)^2\,,
\label{d5metres3}
\ee
where
\be
\Omega\equiv X^3\, c^2 + s^2\, \cosh^2\vp\,.
\ee
This expression reduces to the one found in \cite{d5gauge} if
$\varphi=0$.  In that case, the scalar $X$ parameterises deformations
of $S^5$ corresponding to inhomogeneities of codimension 1 of the
foliation by $S^3\times S^1$.  When the pseudoscalar $\varphi$ is
included too, the inhomogeneities remain of codimension 1, but with a
slightly more complicated structure.  In addition, there is a sort of
``twist'' in the $S^3\times S^1$ product structure of the homogeneous 
foliating surfaces, as indicated by the cross-term between the
interval $d\phi_3$ on $S^1$, and the 1-form $\sigma_3$ on $S^3$.

    Finally, substituting our result for the internal hatted metric
$d\hat s_5^2$ into (\ref{d10metans}), we arrive at the conjectured
ten-dimensional metric Ansatz for this two-scalar truncation:
\bea
d\hat s_{10}^2 &=& \Delta^{\fft12}\, \cosh\vp\, ds_5^2 + \fft{X\,
\Delta^{\fft12}}{\cosh\vp}\, d\xi^2 + \fft{c^2\,
X^{-1}}{4\Delta^{\fft12}\, \cosh\vp}\, (\sigma_1^2+\sigma_2^2)\nn\\
&&+ \fft{c^2\, X\, \Delta^{\fft12}\, \cosh\vp}{4\Omega}\, \sigma_3^2 +
\fft{s^2\,\Omega}{\Delta^{\fft32}\, \cosh\vp}\, \Big(d\phi_3 - \fft{c^2\,
\sinh^2\vp}{2\Omega}\, \sigma_3\Big)^2\,.\label{d10metric}
\eea

\subsection{The field-strength Ans\"atze}

    There does not seem to be any straightforward way to
determine the Ansatz for the Kaluza-Klein reduction other fields of
the ten-dimensional type IIB theory, in this two-scalar reduction.  We
know that when $\vp$ is taken to be zero, the Ansatz must reduce to
one encompassed by the results in \cite{d5gauge}.  In particular, the
remaining scalar field $X$ enters in the Ansatz for the self-dual
5-form, whilst the dilaton, axion and 3-form field strengths of the
type IIB theory vanish when $\vp=0$.  Since it is a pseudoscalar, the
field $\vp$ enters at the linearised level in the Ansatz for the
NS-NS and R-R 2-form potentials $\hat A_\2 \equiv \hat A_\2^{\rm NS}$
and $\hat A_\2^{\rm RR}$ \cite{krvann}.  

   The relevant bosonic equations of motion of the type IIB theory are
\bea
\hat R_{MN} &=&\ft1{96} \hat H^2_{MN}
+ \ft14  \Big((\hat{F}^{1}_\3)^2_{MN} -
\ft1{12}(\hat{F}^1_\3)^2 \hat{g}_{MN}\Big)
+ \ft14
\Big((\hat{F}^{2}_\3)^2_{MN} - \ft1{12}
(\hat{F}^2_\3)^2\hat{g}_{MN}\Big)\,,\nn\\
d{\hat *\hat F_\3} &=& -\im\, \hat H_\5\wedge \hat F_\3\,,\label{d10eom}\\
d\hat H_\5 &=& -\ft{\im}{2}\, \hat F_\3\wedge \hat {\bar F_\3}\,,\qquad
\hat H_\5 = {\hat *\hat H_\5}\,,\nn
\eea
where we have introduced the notation that
\be
\hat A_\2 \equiv  \hat A_\2^{\rm NS} + \im\, \hat A_\2^{\rm RR}\,.
\ee
We are assuming here that the dilaton and axion of the type IIB theory
vanish in the reduction.  For this to be consistent with the type IIB
equations of motion, it is necessary that
\be
{\hat *\hat F_\3}\wedge \hat{\bar F_\3}=0\,,\qquad 
{\hat *\hat F_\3}\wedge \hat F_\3 = {\hat *\hat {\bar F_\3}}\wedge 
\hat {\bar F_\3}\,.
\ee
We shall restrict our discussion from now on to the linearised level.  

    In the notation that we are using here, the linearised Ansatz for
pseudoscalars $\vp$ will be of the form
\be
\hat A_\2 = \vp\, Y_\2\,,
\ee
where $Y_\2$ is a complex 2-form spherical harmonic satisfying 
\be
d{*Y_\2} = \im\, \lambda\, Y_\2
\ee
on the unit round 5-sphere.  The Ansatz for the self-dual 5-form $\hat
H_\5\equiv \hat G_\5 + {\hat *}\hat G_\5$ includes a Freunnd-Rubin
term $\hat G_\5 = 4\ep_\5$ (we have set the gauge coupling $g=1$
here). Substituting into the type IIB equations of motion, one finds
that the pseudoscalar $\vp$ satisfies the linearised equation of motion
\be
[d{*d\varphi} + \lambda(\lambda-4)\, \varphi\,\ep_\5
]\wedge{*Y_\2}=0\,.
\ee
A 2-form harmonic with eigenvalue $\lambda$ gives a pseudoscalar
$\varphi$ with $m^2 = \lambda(\lambda-4)$.   We want the mass for the
$10$ and $\overline{10}$ members of the massless multiplet, namely $m^2=-3$,  
which therefore requires $\lambda=1$ or $\lambda=3$.  In fact, the required
harmonics are those with $\lambda=3$ (there are none with
$\lambda=1$).  

    There are ten such harmonics on $S^5$, which can be
written in terms of the Killing spinors.  There are Killing spinors $\eta_\pm$
satisfying $D_a\eta_\pm = \pm\, \ft{\im}{2}\, \Gamma_a\, \eta_\pm$.
It turns out that the required 2-form harmonics are given by the
construction
\be
Y_{ab} = \bar\eta_-\, \Gamma_{ab}\, \eta_+\,,\label{yfromks}
\ee
where $\eta_-$ and $\eta_+$ are any two Killing spinors of the minus
and plus kinds respectively.  Solving for the Killing spinors, and
substituting into (\ref{yfromks}), we find
that one of the ten harmonics has a structure that is particularly
naturally adapted to our parameterisation of the sphere, namely
\be
Y_\2 = e^{\im\, \phi_3}\, \Big( c\, d\xi\wedge \sigma_3 + \ft12 s\,
c^2\, \sigma_1\wedge \sigma_2 - \im\, s\, c^2\, \sigma_3\wedge d\phi_3
\Big)\,.
\ee
One may expect that this harmonic, or a closely related construction, 
will play a significant r\^ole in
the construction of the reduction Ansatz at the full non-linear order,
but we have not yet completed this investigation.

\subsection{Oxidation of five-dimensional solutions}
   
     Given the conjectured metric reduction Ansatz, we can oxidise the
metric in any solution of the two-scalar truncation of
five-dimensional maximal gauged supergravity back to a solution of
type IIB supergravity in $D=10$.  In principle, one can solve the
equations of motion in this two-scalar sector to obtain a
supersymmetric domain wall solution, which has an interpretation as
the RG-flow equations on the strongly coupled field theory side, as
discussed in \cite{fgpw}.  Unfortunately the equations seem not to
allow an explicit solution in terms of elementary functions.

   One simple oxidation that we {\it can} perform is to take the
$D=5$ solution corresponding to the second (non-trivial) supersymmetric 
stationary point of the potential.  This corresponds to the stationary
point of (\ref{d5pot2}) with \cite{kpw}
\be
X= 2^{-\fft13}\,,\qquad \sinh\vp = \ft1{\sqrt3}\,.
\ee
(The fully-supersymmetric stationary point is at $X=1$, $\vp=0$.)
Substituting into (\ref{d5metres3}), we find that the internal
5-sphere metric $d\hat s_5^2$ at this stationary point is given by
\be
d\hat s_5^2 = \fft{3}{2^{7/3}}\, \Big[d\xi^2 + \fft{c^2}{2(1+s^2)}\, 
(\sigma_1^2+\sigma_2^2) 
+ \fft{2 c^2}{3+ 5 s^2}\, \sigma_3^2
+ \fft{s^2\, (3+5s^2)}{3(1+s^2)^2}\,  
\Big(d\phi_3 - \fft{c^2}{3+5s^2}\, \sigma_3\Big)^2\Big]\,.
\label{d5metsol}
\ee

\section{Conclusion}

     In this paper, we have obtained the metric Ans\"atze for two
examples of Kaluza-Klein sphere reductions, both of which involve
pseudoscalar as well as scalar fields.  The first example is the $S^7$
reduction of eleven-dimensional supergravity, with a truncation from
$N=8$ to the $N=2$ theory with $U(1)^4$ gauge fields, three dilatons
and three axions.  Among other uses this reduction allows one to study
the eleven-dimensional geometries corresponding to the lifting of
the four-dimensional BPS AdS black hole and domain-wall solutions
\cite{duliu} of gauged supergravity.  Our results generalise those
obtained previously in \cite{ten}, where the problem was studied in the
absence of the three axionic scalars.  

   Our second example is a truncation of five-dimensional maximal
gauged supergravity, to a subsector in which two spin-0 fields are
retained, one of which is a scalar, and the other a pseudoscalar.
This truncation retains the fields necessary for describing a second
supersymmetric vacuum in $D=5$, with $N=2$ supersymmetry and
$SU(2)\times U(1)$ invariance, in addition to the
maximally-supersymmetric one with $SO(6)$ invariance \cite{kpw}.  The
metric reduction Ansatz that we obtain here allows one to study the
ten-dimensional geometries corresponding to the lifting of solutions
of the five-dimensional theory.  In principle, this can include the
renormalisation-group flow \cite{fgpw} associated with the second
supersymmetric extremum, although the explicit form of this
five-dimensional solution is not known.

\section*{Acknowledgement}
 
    We are grateful to Jim Liu, Krzysztof Pilch, Tuan Tran and Nick Warner
for discussions.

\appendix
\section*{Appendix A}

    In this appendix, we present the explicit form of the $SO(8)$
triality rotation that we used in section 2.1 in order to simplify the 
Kaluza-Klein metric reduction Ansatz:
\bea
&&K_{12}\longrightarrow \ft12(K_{12} + K_{34} + K_{56} + K_{78})\,, \qquad
K_{13}\longrightarrow \ft12(K_{13} - K_{24} + K_{57} - K_{68})\,,\nn\\
&&K_{14}\longrightarrow \ft12(K_{14} + K_{23} + K_{58} + K_{67})\,, \qquad
K_{15}\longrightarrow \ft12(K_{15} - K_{26} + K_{37} - K_{48})\,,\nn\\
&&K_{16}\longrightarrow \ft12(K_{16} + K_{25} + K_{38} + K_{47})\,, \qquad
K_{17}\longrightarrow \ft12(K_{17} - K_{28} + K_{35} - K_{46})\,,\nn\\
&&K_{18}\longrightarrow \ft12(K_{18} + K_{27} + K_{36} + K_{45})\,, \qquad
K_{23}\longrightarrow \ft12(K_{23} + K_{14} - K_{58} - K_{67})\,,\nn\\
&&K_{24}\longrightarrow \ft12(K_{24} - K_{13} + K_{57} - K_{68})\,, \qquad
K_{25}\longrightarrow \ft12(K_{25} + K_{16} - K_{38} - K_{47})\,,\nn\\
&&K_{26}\longrightarrow \ft12(K_{26} - K_{15} + K_{37} - K_{48})\,, \qquad
K_{27}\longrightarrow \ft12(K_{27} + K_{18} - K_{36} - K_{45})\,,\nn\\
&&K_{28}\longrightarrow \ft12(K_{28} - K_{17} + K_{35} - K_{46})\,, \qquad
K_{34}\longrightarrow \ft12(K_{34} + K_{12} - K_{56} - K_{78})\,,\nn\\
&&K_{35}\longrightarrow \ft12(K_{35} + K_{17} + K_{28} + K_{46})\,, \qquad
K_{36}\longrightarrow \ft12(K_{36} + K_{18} - K_{27} - K_{45})\,,\nn\\
&&K_{37}\longrightarrow \ft12(K_{37} + K_{15} + K_{26} + K_{48})\,, \qquad
K_{38}\longrightarrow \ft12(K_{38} + K_{16} - K_{25} - K_{47})\,,\nn\\
&&K_{45}\longrightarrow \ft12(K_{45} + K_{18} - K_{27} - K_{36})\,, \qquad
K_{46}\longrightarrow \ft12(K_{46} + K_{35} - K_{17} - K_{28})\,,\nn\\
&&K_{47}\longrightarrow \ft12(K_{47} + K_{16} - K_{25} - K_{38})\,, \qquad
K_{48}\longrightarrow \ft12(K_{48} + K_{37} - K_{15} - K_{26})\,,\nn\\
&&K_{56}\longrightarrow \ft12(K_{56} + K_{12} - K_{34} - K_{78})\,, \qquad
K_{57}\longrightarrow \ft12(K_{57} + K_{13} + K_{24} + K_{68})\,,\nn\\
&&K_{58}\longrightarrow \ft12(K_{58} + K_{14} - K_{23} - K_{67})\,, \qquad
K_{67}\longrightarrow \ft12(K_{67} + K_{14} - K_{23} - K_{58})\,,\nn\\
&&K_{68}\longrightarrow \ft12(K_{68} + K_{57} - K_{13} - K_{24})\,, \qquad
K_{78}\longrightarrow \ft12(K_{78} + K_{12} - K_{34} - K_{56})\,.
\eea

\section*{Appendix B}

    In this Appendix, we collect some results on the geometry of the
7-sphere. We can describe $S^7$ as the unit sphere in $R^8$, with 8
real coordinates $x_I$;
\be
x_I\, x_I = 1\,.
\ee
As such, it has a manifest $SO(8)$ symmetry, with 28 Killing vectors
$K_{IJ}$ given by
\be
K_{IJ} = x^I\, \fft{\del}{\del x_J} -  x^J\, \fft{\del}{\del x_I}\,.
\label{killing}
\ee
We can also describe $S^7$ as the unit sphere in $C^4$, with 4 complex 
coordinates $z_i$:
\be
\bar z_i\, z_i = 1\,.\label{bzz}
\ee
We can relate these complex coordinates to the previous real ones as
follows:
\be
z_1 = x_1 + \im\, x_2\,,\quad 
z_2 = x_3 + \im\, x_4\,,\quad
z_3 = x_5 + \im\, x_6\,,\quad
z_4 = x_7 + \im\, x_8\,.
\ee
We can parameterise these complex coordinates as
\be
z_1 = \mu_1\, e^{\im\, \phi_1}\,,\quad 
z_2 = \mu_2\, e^{\im\, \phi_2}\,,\quad
z_3 = \mu_3\, e^{\im\, \phi_3}\,,\quad
z_4 = \mu_4\, e^{\im\, \phi_4}\,,
\ee
where (\ref{bzz}) implies that
\be
\sum_{i=1}^4 \mu_i^2 =1\,.
\ee
These $(\mu_i, \phi_i)$ coordinates are precisely the ones used for
describing higher-dimensional rotating black holes in \cite{mype},
and in the $S^7$ reduction Ansatz obtained in \cite{ten}.

   From the coordinate transformations above, it is straightforward to
establish that the real derivatives $\del/\del x_I$ that appear in the
Killing vectors (\ref{killing}) are given by
\be
\fft{\del}{\del x_1} = \cos\phi_1\, \fft{\del}{\del\mu_1} -
\fft{\sin\phi_1}{\mu_1}\, \fft{\del}{\del\phi_1}\,,\qquad
\fft{\del}{\del x_2} = \sin\phi_1\, \fft{\del}{\del\mu_1} +
\fft{\cos\phi_1}{\mu_1}\, \fft{\del}{\del\phi_1}\,,
\ee
with analogous expressions involving $(\mu_2,\phi_2)$,
$(\mu_3,\phi_3)$ and $(\mu_4,\phi_4)$ for the pairs $(x_3,x_4)$,
$(x_5,x_6)$ and $(x_7,x_8)$ respectively.  It is easy to see from this
that the four Killing vectors $K_{12}$, $K_{34}$, $K_{56}$ and
$K_{78}$ are simply of the form:
\be
K_{12}= \fft{\del}{\del\phi_1}\,,\quad
K_{34}= \fft{\del}{\del\phi_2}\,,\quad
K_{56}= \fft{\del}{\del\phi_3}\,,\quad
K_{78}= \fft{\del}{\del\phi_4}\,.
\ee
These are the four commuting $U(1)$ generators.  It is convenient to
write them as $\del_{\phi_1}$, {\it etc}.

    We also note that the Killing-vector bilinears in the top 3
lines in (\ref{inversemet}) are also relatively simple, when expressed 
in terms of the $\mu_i$ and $\phi_i$ coordinates.  After some algebra
we find, for example, that
\be
K_{13}^2 + K_{14}^2 + K_{23}^2 + K_{24}^2=
(\mu_1\, \del_{\mu_2} - \mu_2\,\del_{\mu_1} )^2 + 
\fft{\mu_2^2}{\mu_1^2}\, \del_{\phi_1}^2 + 
\fft{\mu_1^2}{\mu_2^2}\, \del_{\phi_2}^2 
\ee
with analogous results for the other five combinations.


\begin{thebibliography}{99}

\bibitem{dwn} B. de Wit and H. Nicolai, {\sl The consistency
of the $S^{7}$ truncation in $D=11$ supergravity}, Nucl. Phys. B281
(1987) 211.

\bm{ten} M. Cveti\v{c}, M.J. Duff, P. Hoxha, J.T. Liu, H. L\"u,
J.X. Lu, R. Martinez-Acosta, C.N. Pope, H. Sati and T.A. Tran, {\sl
Embedding AdS black holes in ten and eleven dimensions},
Nucl. Phys. {\bf B558} (1999) 96, hep-th/9903214.

\bm{d4gauge} M. Cveti\v c, H, L\"u and C.N. Pope, {\sl
Four-dimensional $N=4$, $SO(4)$ gauged supergravity from $D=11$},
hep-th/9910252, to appear in Nucl. Phys. {\bf B}.

\bm{kpw} A. Khavaev, K. Pilch and N.P. Warner, {\sl New vacua of
gauged $N=8$ supergravity in five dimensions}, hep-th/9812035.

\bm{clps} M. Cveti\v c, H. L\"u, C.N. Pope and A. Sadrzadeh, {\sl
Consistency of Kaluza-Klein sphere reductions of symmetric
potentials}, hep-th/0002056.

\bm{vann1} H. Nastase, D. Vaman and P. van Nieuwenhuizen, {\sl
Consistent nonlinear KK reduction of 11-D supergravity on AdS$_7\times
S^4$ and selfduality in odd dimensions}, Phys. Lett. {\bf B469} (1999)
96, hep-th/9905075.

\bm{vann2} H. Nastase, D. Vaman and P. van Nieuwenhuizen, {\sl
Consistency of the AdS$_7\times S_4$ reduction and the origin of
self-duality in odd dimensions}, hep-th/9911238.  

\bm{d6gauge} M. Cveti\v{c}, H. L\"u and C.N. Pope, {\sl Gauged
six-dimensional supergravity from massive type IIA},
Phys. Rev. Lett. {\bf 83} (1999) 5226,  hep-th/9906221. 

\bm{nasvam} H. Nastase and D. Vaman, {\sl On the nonlinear KK reductions
on spheres of supergravity theories}, hep-th/0002028.  

\bibitem{dwn00}
B. de Wit and H. Nicolai,
{\sl $N=8$ supergravity with local $SO(8) \times SU(8)$ invariance},
Phys. Lett. {\bf B108} (1982) 285.

\bibitem{dwn0}
B. de Wit and H. Nicolai,
{\sl N=8 supergravity},
Nucl. Phys. {\bf B208} (1982) 323.

\bm{dwnw} B. de Wit, H. Nicolai and N.P. Warner, {\sl The embedding of
gauged $N=8$ supergravity into $D=11$ supergravity}, Nucl. Phys. {\bf
B255} (1985) 29.

\bibitem{duliu}
M.J. Duff and J.T. Liu,
{\sl Anti-de Sitter black holes in gauged N=8 supergravity},
Nucl. Phys. {\bf B554} (1999) 237, hep-th/9901149.

\bm{dist} M. Cveti\v c, S. Gubser H. L\"u and C.N. Pope,
{\sl Symmetric potentials of gauged supergravities in diverse
dimensions and Coulomb branch of gauge theories}, hep-th/9909121, to
appear in Phys. Rev. {\bf D}.

\bm{KLT} P. Kraus, F. Larsen and S.P. Trivedi, {\sl The Coulomb branch
of gauge theory from rotating branes}, JHEP {\bf 9903} (1999) 003,
hep-th/9811120.

\bibitem{FGPW} D.Z. Freedman, S.S. Gubser, K. Pilch and N.P. Warner,
{\sl Continuous distributions of D3-branes and gauged supergravity},
hep-th/9906194.

\bibitem{BS} A. Brandhuber and K. Sfetsos, {\sl Nonstandard
compactification with mass gaps and Newton's Law}, hep-th/9908116.

\bibitem{BSI} I. Bakas and K. Sfetsos, {\sl States and curves of
five-dimensional gauged supergravity}, hep-th/9909041.

\bm{BBS} I. Bakas, A. Brandhuber and K. Sfetsos, {\sl Domain walls of
gauged supergravity, M-branes and algebraic curves}, hep-th/9912132.

\bm{grw} M. Gunaydin, L.J. Romans and N.P. Warner, {\sl Gauged
$N=8$ supergravity in five dimensions}, Phys. Lett. {\bf B154} (1985)
268.

\bm{fgpw} D.Z. Freedman, S.S. Gubser, K. Pilch and N.P. Warner, {\sl 
Renormalization group flows from holography, supersymmetry and a
C-theorem}, hep-th/9904017.

\bm{d5gauge} H. L\"u, C.N. Pope and T.A. Tran, {\sl Five-dimensional 
$N=4$, $SU(2)\times U(1)$ gauged supergravity from type IIB},
Phys. Lett. {\bf B475} (2000) 261, hep-th/9909203.

\bm{krvann} H.J. Kim, L.J. Romans, and P. van Nieuwenhuizen, {\sl Mass
spectrum of ten dimensional $N=2$ supergravity on $S^5$},
Phys. Rev. {\bf D32} (1985) 389.

\bibitem{mype}
R.C. Myers and M.J. Perry,
{\sl Black holes in higher dimensional spacetimes},
Ann. Phys. {\bf 172} (1986) 304.

\end{thebibliography}
\end{document}